\documentclass[useAMS,usenatbib]{mn2e}
\usepackage{graphicx}
\usepackage{subfigure}
\usepackage{amsmath}
\usepackage{amssymb}
\usepackage{color}
\usepackage{natbib}
\usepackage{slantsc}

\title[]{The Variation in Molecular Gas Depletion Time among Nearby Galaxies:
What are the Main Parameter Dependencies?}
\author[M.-L. Huang et al.]{Mei-Ling Huang$^{1}$\thanks{E-mail:
mlhuang@mpa-garching.mpg.de}, Guinevere Kauffmann$^{1}$\\
$^{1}$Max-Planck Institute for Astrophysics, Karl-Schwarzschild-Str. 1, D-85748 Garching, Germany\\
}

\begin{document}
\date{ in original form 2014 Jan}
\pagerange{\pageref{firstpage}--\pageref{lastpage}} \pubyear{2002}
\maketitle
\label{firstpage}

\begin{abstract}
We re-analyze correlations between global molecular gas depletion time
($t_{dep}$) and galaxy parameters for nearby galaxies from the COLD GASS 
survey. We improve on previous work of \citet{sanb} 
by estimating star formation rates (SFRs) using the combination of GALEX FUV 
and WISE 22 $\micron$ data and by deriving $t_{dep}$ within a fixed 
aperture set by the beam size of gas observation.  
In our new study we find correlations with much smaller scatter.
Dependences of the depletion time on galaxy structural parameters such as
stellar surface density and concentration index are now weak or absent.
We demonstrate that the {\em primary} global parameter correlation is 
between t$_{dep}$ and sSFR; all other remaining correlations can 
be shown to be induced by this primary dependence.  
This implies that galaxies with high current-to-past-averaged
star formation activity, will drain their molecular gas reservoir sooner.
We then analyze $t_{dep}$ on 1-kpc scales in galactic
disks using data from the HERACLES survey. There is remarkably good
agreement between the global t$_{dep}$--sSFR relation for the COLD
GASS galaxies and that derived for 1 kpc scale grids {\em in disks}.
This leads to the conclusion that the local molecular gas
depletion time in galactic disks is dependent on the local fraction of
young-to-old stars.
\end{abstract}


\section{Introduction} 

Numerous studies have attempted  to characterize the relation between 
star formation rate surface density ($\rm \Sigma_{SFR}$) and total 
gas surface density ($\rm \Sigma_{gas}$). Such relations are often referred to as  
Kennicutt-Schmidt ``laws''\citep{ken98}.
The Kennicutt-Schmidt ``law'' is usually expressed as 
\[
\Sigma_{SFR} \propto \Sigma_{gas}^{N},
\]
where $\rm \Sigma_{SFR}$ and $\rm \Sigma_{gas}$ are in units of 
M$_{\sun}$ yr$^{-1}$ kpc$^{-2}$ and M$_{\sun}$ pc$^{-2}$.

\citet{ken98} found N $\sim$1.4 on global scales for normal and starburst 
galaxies using H$\alpha$ as a star formation rate (SFR) indicator and CO 
and H\textsc{i} line emission as tracers of total cold gas content.
It has been demonstrated that this value of N  can be explained by 
gravitational instability: if stars are formed constantly each free-fall 
time and gas scale height is assumed to be constant, N is equal to 1.5 
(e.g., \citealt{mad}; \citealt{elm}).
Other studies have found values for N that vary in the range between 1 and 3  
(see review by \citealt{elm11}).
The scatter between different studies is likely due to the fact that 
a variety of galaxy samples and
SFR tracers were used,  and that the relation was evaluated over a range of
different spatial scales.

Stars form directly in molecular clouds, so the molecular gas,
rather than the total or atomic gas, is believed  to be linked 
more directly to the star formation.
\citet{won} found a roughly linear correlation
between $\rm \Sigma_{SFR}$ and $\rm \Sigma_{H2}$ using  H$\alpha$ 
and CO observations of 7 gas-rich spirals. In more recent studies, 
\citet{big08} explored the correlation between 
$\rm \Sigma_{SFR}$ on sub-kpc scales estimated from FUV plus 24 $\micron$ 
fluxes and $\rm \Sigma_{H2}$ derived from CO (J=2-1) line luminosities 
using a sample of 18 nearby spiral galaxies from the HERACLES 
project \citep{ler08}. They reported a constant star formation efficiency with 
depletion time $\sim$ 2 Gyr among these spirals, independent of local 
conditions such as  orbital timescale, midplane gas pressure and 
disk stability \citep{ler08}.
Their results appeared to  suggest that star formation is likely 
to be a localised process linked only to the local quantity of molecular gas.
However, in subsequent work,  \citet{mom} found a super-linear Kennicutt-Schmidt law 
at sub-kpc resolution with slope N = 1.3 -- 1.8, based on CO (J=1-0) plus  
H$\alpha$ and 24 $\micron$ data for 10 nearby spiral galaxies.
These results would  argue that  the star formation efficiency in disks
depends on local gas surface density. 
Differences with the published HERACLES results were attributed to the fact that   
CO(J = 2-1) was used as a tracer of molecular gas mass. 
If CO(J = 2-1) is enhanced in  regions with higher star formation 
(and hence higher excitation), CO(J = 2-1) may not be a linear tracer of the 
underlying molecular gas mass.

The COLD GASS project investigated the relation between molecular gas 
depletion time (t$_{dep}$) and global galaxy parameters  for a 
representative sample of $\sim$300 galaxies with stellar masses 
$\rm 10^{10}-10^{11.5} M_{\sun}$, at redshift 0.02 -- 0.05 \citep{sana}.
CO(1-0) line measurements were combined with ancillary optical and FUV data 
from the Sloan Digital Sky Survey (SDSS) and the Galaxy Evolution Explorer (GALEX) 
satellite to carry out a statistical analysis of global relations between  
star formation and molecular gas for nearby galaxies \citep{sanb}. The global 
star formation efficiency (SFE) or molecular gas depletion time was found to 
be dependent on a variety of galaxy parameters, including stellar mass, stellar 
surface mass density, concentration of the light (i.e. bulge-to-disk ratio),
NUV$-r$ colour and specific star formation rate. The strongest dependences were on
colour and specific star formation rate.

In subsequent work, \citet{ler13} studied the global Kennicutt-Schmidt law for 30 
nearby spirals using CO(2-1) line luminosities as the molecular gas tracer and H$\alpha$ plus 
24$\mu$m luminosities as the SFR tracer.
They found  that all dependences of t$_{dep}$ on global 
galaxy properties vanished after applying a  CO-to-H$_2$ conversion factor 
($\rm \alpha_{CO}$) that depended on dust-to-gas-ratio.
\citet{sanb} tested whether there was any relation between  t$_{dep}$ and
gas-phase metallicity determined from nebular emission lines in their sample and found 
a null result. Since dust-to-gas ratio and metallicity are well-correlated, this
would argue that the explanation given in \citet{ler13} for apparent variations
in molecular gas depletion time, cannot be the explanation for the results presented
in \citet{sanb}.

We note that the sample of galaxies studied by \citet{ler13} spans a much
narrower range in specific  star formation rate than the COLD GASS sample.
In particular, massive, bulge-dominated galaxies with low gas fractions and low SFR/$M_*$
are absent. These are the galaxies that contribute most to the observed
variations on depletion time (see Figure 7 in \citealt{sanb}).    

One hypothesis that has not yet been explored is that different SFR tracers may be the
origin of the differing conclusions of \citet{ler13} and \citet{sanb}.
\citet{sanb} used spectral energy distribution (SED) fitting technique to derive 
SFR by fitting UV and optical broad-band photometry to a grid of SED models.
Correction for dust extinction was not done directly, but by adopting a set
of priors that linked the range of possible values of A$_V$ in the
model grid with the NUV-r colours and 4000 \AA\ break strengths of the galaxies. 
These priors were calibrated using results from the GMACS (Galaxy Multiwavelength 
Atlas From Combined Surveys) sample \citep{jon}
, for which accurate star formation rates and extinction measurements 
could be obtained using UV and far-IR data. 

In this paper, we re-derive SFRs for galaxies in the COLD GASS sample
by combining GALEX FUV and WISE 22\micron\ data (Wide-field Infrared Survey Explorer;\citealt{wri}). 
We thus eliminate the need for statistical dust correction methods.
We also improve on the \citet{sanb} analysis by measuring molecular gas masses
and SFR surface densities within apertures that are exactly matched in radius.  
We show that not only does this reduce the scatter in our previously derived relations between molecular
gas depletion time and galaxy parameters, but also changes some of the scalings. Correlations with
galaxy structural parameters such as stellar surface mass density and the concentration index
of the light are considerably weakened or entirely absent. We are able to show that the only
galaxy parameter that is significantly correlated with the global depletion time of the
molecular gas is the ratio of young-to-old
stars in the galaxy. We also utilize molecular gas and star formation maps from the HERACLES
survey to demonstrate that very similar results are found on sub-kpc scales.

\section{Data}    

\subsection{COLD GASS} 

\subsubsection{Molecular gas and galaxy parameters}
Our data set is drawn from the COLD GASS survey catalogue 
\citep{sana, sanb,sanc}, which is a subsample of the GASS 
survey \citep{cat}. COLD GASS contains CO ($J=1-0$) line measurements 
from the IRAM 30m telescope for $\sim$360 nearby galaxies with 
stellar masses in the range  $10^{10}-10^{11.5} M_{\sun}$ 
and redshifts in the range $0.025 < z < 0.05$.
The reader is referred to \citet{sana} for 
a detailed description of the sample selection and the observations. 

The COLD GASS catalogue also includes global galaxy parameters such as
stellar mass, stellar surface mass density, concentration index 
and NUV-r colour for each targeted galaxy. The first three parameters were  
taken from the MPA/JHU value-added catalogs 
(http://www.mpa-garching.mpg.de/SDSS).
Stellar masses were calculated by fitting SDSS 5-band optical magnitudes  
to stellar population synthesis model grids. Stellar surface mass 
density is defined as $M_{*}/(2\pi R^{2}_{50,z})$, where 
$R_{50,z}$ is  the radius containing 50\% of the Petrosian flux 
in the $z$-band. Concentration index is  defined as $R_{90}/R_{50}$ where
$R_{90}$ and $R_{50}$ are the radii enclosing 90\% and 50\% of the 
total $r$-band light. The concentration index is well-correlated 
with the bulge-to-total luminosity ratio of the galaxy (\citealt{gad}; \citealt{wei}).
The NUV-r colours of the galaxies were measured using SDSS $r$-band 
and GALEX NUV images, which were corrected for Galactic extinction 
following \citet{wyd}.  

\subsubsection{GALEX FUV and WISE 22$\micron$ Data}
Instead of adopting the catalog SFRs derived using SED-fitting methods,  
we derive new SFRs using the combination of FUV emission from 
GALEX data and mid-infrared 22$\micron$ emission from WISE data.
FUV emission traces recent unobscured star formation in the galaxy. 
The missing part of the emission from young, massive stars obscured by surrounding dust 
can be recovered via the measured IR emission, which originates from the dust that is heated 
by absorbed FUV light.

GALEX provides FUV images with effective wavelength at 1528 \AA\  and 
angular resolution $\sim$4.3 arcsec FWHM. We draw the FUV maps from 
GALEX Data Release 7 products. When several maps from different surveys 
are available, we select the maps with the longest exposure time. 
Most of our FUV images are from the Medium Imaging survey (MIS) with 
typical exposure time $\sim$1500 seconds;  81 out 
366 galaxies only have data from the All-sky Imaging survey (AIS) with typical exposure 
time $\sim$100 -- 200 seconds. {\bf The limiting flux for 
MIS is $\sim$1.4 $\mu$Jy and for AIS, it is $\sim$0.2 $\mu$Jy.}
WISE provides  22$\micron$ images of the 
whole sky, with angular resolution 12 arcsec and  5$\sigma$ 
point-source sensitivity $\sim$6mJy.
With both GALEX and WISE all-sky maps available, we are able to 
derive more accurate  SFRs for the COLD GASS sample galaxies.

\begin{figure} 
\begin{center}
 \includegraphics[scale=0.38]{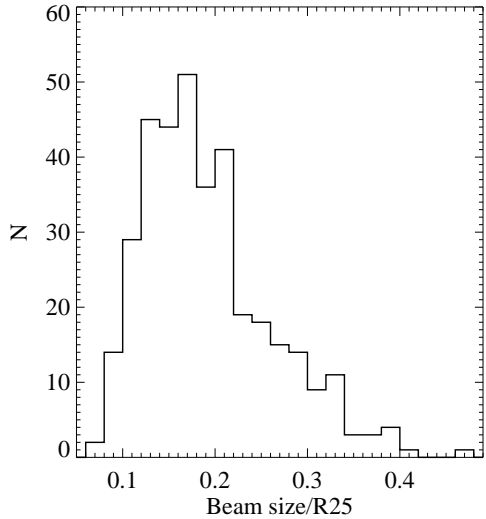}
    \caption{Ratio of IRAM beamsize to R$_{25}$ for galaxies in the  COLD GASS sample.}
  \label{f1}
\end{center}
\end{figure}

\subsection{HERACLES}  

To compare molecular gas depletion time trends  on sub-kpc scales with
results obtained globally,  
we utilize publicly available data from the HERA 
CO-Line Extragalactic Survey (HERACLES; \citealt{ler08}).
HERACLES has released CO ($J=2-1$) maps for 48 nearby galaxies,
achieving a spatial resolution $\sim$13 arcsec and an 
average H$_{2}$ surface density detection limit of $\sim$3 M$_{\sun}$pc$^{-2}$.
For our comparisons with the COLD GASS results , we select 20 massive 
galaxies with log (M$_*$/M$_{\sun}$) $>$ 10 from the catalog 
in \citet{ler13}; these selected galaxies are all located within a distance 
of $\sim$ 20 Mpc.

A variety of ancillary data is available for the galaxies we select from HERACLES sample.   
This includes FUV images from GALEX AIS (four galaxies) and 
Nearby Galaxy Survey (NGS; \citealt{gil}), 24 $\micron$ data from the 
Spitzer Infrared Nearby Galaxies Survey (SINGS; \citealt{ken03}),
H\textsc{i} maps from The H\textsc{i} Nearby Galaxies Survey 
(THINGS; \citealt{wal}). SINGS provides MIPS 24$\mu$m images 
with an angular resolution $\sim$6 arcsec
and 3 $\sigma$ sensitivity $\sim$0.21 MJy sr$^{-1}$. 
The maps (natural-weighting) from THINGS have an angular resolution 
of $\sim$11 arcsec and are sensitive to $\rm \Sigma_{HI}$ 
$\geq$ 0.5 M$_{\sun}$ pc$^{-2}$.

The combination of these public datasets enables us to derive   
gas and SFR surface densities as a function of position in the galaxy
by dividing the galaxy into a set of square
cells using a 1 kpc$^2$ grid. Galaxy parameters such as distance, R$_{25}$,
inclination angle, position angle are taken from \citet{ler13}.
Stellar mass contained within each grid cell is estimated by fitting 
SDSS 5-band photometry measured for the cell  to stellar population models (see sec 3.2).

\begin{figure} 
\begin{center}
 \includegraphics[scale=0.4]{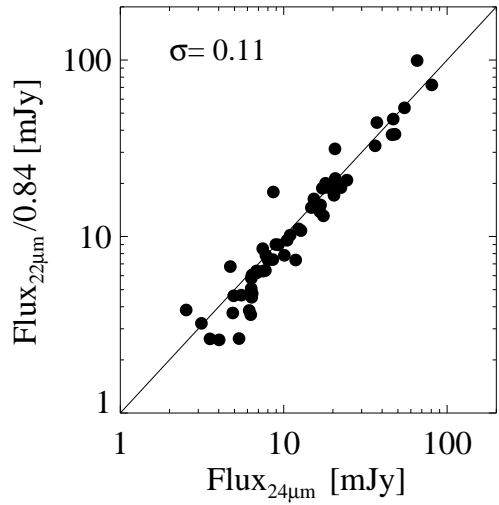}
    \caption{Scaled WISE 22\micron\ fluxes are plotted against Spitzer MIPS 24\micron\ fluxes for
galaxies with WISE detections in the GMACS catalog.}
  \label{f2}
\end{center}
\end{figure}

\begin{figure} 
\begin{center}
 \includegraphics[scale=0.36]{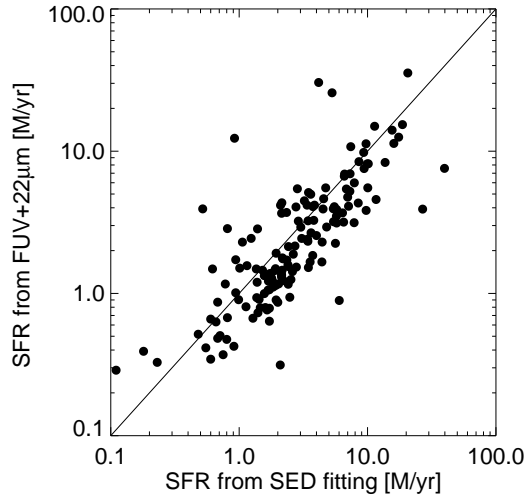}
    \caption{Comparison of SFR estimates from the SED-fitting method presented 
in Saintonge et al (2011) and the combination of FUV plus 22$\mu$m.}
  \label{f3}
\end{center}
\end{figure}

\section{Analyses}  

\subsection{Molecular gas depletion time in the COLD GASS sample}  
Since the IRAM 22$\arcsec$\ beam size does not cover the entire optical disk of the 
galaxies in the COLD GASS sample, aperture corrections were applied to derive total 
molecular gas masses in \citet{sana}. In Fig.\ref{f1}, we plot the ratio of the IRAM 
beam size to R$_{25}$, the 25 mag arcsec$^{-2}$ $g$-band isophotal radius.
As can be seen, the beam radius is typically only a quarter of the optical radius. 
Because the molecular gas is generally quite concentrated towards the
center of the galaxy, the corrections to total gas mass are not large {\em on average}
(Saintonge et al 2011a).
Nevertheless, when calculating molecular gas depletion time, it is much more
accurate to evaluate all quantities within the central 22$\arcsec$ region
of the galaxy.

We thus use the observed CO fluxes rather than the corrected CO fluxes from the 
COLD GASS survey catalogues to derive our estimates of the molecular gas mass.
We apply a  CO-to-H$_2$ conversion factor, $\rm \alpha_{CO}$, in the 
catalogue to calculate the molecular mass from observed CO line luminosities . In the 
catalogue, $\rm \alpha_{CO}$ is assumed to have
the Galactic value (4.35 M$_{\sun}$(K km s$^{-1}$     
pc$^{2}$)$^{-1}$) for the main sample  galaxies. For the subset of
interacting starburst galaxies with high SFR/$M_*$ values that were added as extra
targets towards the end of the program, $\rm \alpha_{CO}$ is assumed to have  
 the  ``ULIRG'' value (1 M${_{\sun}}$(K km s$^{-1}$ pc$^{2}$)$^{-1}$). 
Note that there are only 11 galaxies that  require the ULIRG-type $\rm \alpha_{CO}$ 
in the catalogue. More detailed discussion  can be found in Sec 2.3 of \citet{sanc}.

To derive the depletion time, SFR must be estimated within the same region where 
molecular gas is measured. We place a Gaussian 
IRAM beam on the FUV and WISE 22 $\micron$ maps to derive the central SFR.
We note that this was not done in the analysis of \citet{sanb}, who used
total SFR estimates and aperture-corrected CO fluxes.
What we want to estimate,  is  how fast 
the molecular gas is consumed into stars. To answer this question as accurately as possible, 
it is clearly preferable that the apertures for the molecular gas and 
SFR measurements should be closely matched.

We first run {\rm SExtractor} on the WISE images and mask the detected sources, with
the exception of the central target.
To remove the background, we first generate the sky image by 
selecting pixels with values $<$ 3 $\sigma$ above the median value
of the image. We then subtract the median value of the sky image 
as background. Following \citet{wri} and \citet{jar}, we apply a 
colour correction and a flux correction for extremely red sources.
Finally, we scale the 22 $\micron$ WISE fluxes so that they
have the same normalization as the MIPS 24$\mu$m fluxes.  
Using the SED template of normal galaxies from \citet{cha}, the flux ratio 
between WISE 22 $\micron$ and Spitzer MIPS 24 $\micron$ filters is $\sim$0.841 
at a redshift $z< 0.05$.

One question is whether there is much scatter around this ratio.
We use the GMACS catalog from \citet{jon} to answer this question. The catalogue 
provides a sample located in two sky regions: the Lockman Hole (part of the 
Spitzer Wide-area InfraRed Extragalactic survey; SWIRE) and the Spitzer 
Extra-Galactic First Look Survey (FLS).  
The median stellar mass  of galaxies in this sample is {\bf5$\times$10$^{10}$ M$_{\sun}$} and
the median redshift is 0.11.
The  22$\micron$ WISE data is not as deep as SWIRE and FLS surveys, so 
we only obtain secure 22$\micron$ measurements for  $\sim$60 galaxies from the catalogue 
of \citet{jon}.  We predict  the fluxes at  MIPS-24$\micron$
by dividing by  0.84, the value estimated from our single template.
The comparison between the pseudo and observed fluxes at Spitzer-24$\micron$  
is plotted in Fig. \ref{f2}. As can be seen, the predicted  values 
(y-axis) match the observed values (x-axis) very well and the 
the scatter is only $\sim$ 0.11 dex.

Our analysis of the FUV images involves first masking stars and background galaxies 
using the same procedure applied to the WISE images. We correct for the effects 
of Galactic extinction using the extinction maps of \citet{sch}. Finally, we adopt 
the combined GALEX FUV plus MIPS 24$\micron$ relation in \citet{ler08} to estimate 
the total SFR. 

In Fig. 3, we compare the total SFRs derived from FUV plus IR with the 
SFRs derived via SED-fitting method from the COLD GASS catalogue. There is no 
significant offset in mean normalization between the two estimates, but there is
considerable scatter of around a factor of two.  There are 9 galaxies for which the 
estimates differ by factors of 5-10. Some of these are very dusty  galaxies 
with  SFR$\rm_{IR}$/SFR$\rm_{UV}$ ratios much larger than the average,  
where  SFRs have been underestimated by the  SED-fitting method. 
The galaxies with SFRs overestimated by SED-fitting method have  
SFR$\rm_{IR}$/SFR$\rm_{UV}$ ratios close to the mean value,  but NUV-r colours 
that are very red, indicative of old stellar populations.

\begin{figure*} 
\begin{center}
 \includegraphics[scale=0.65]{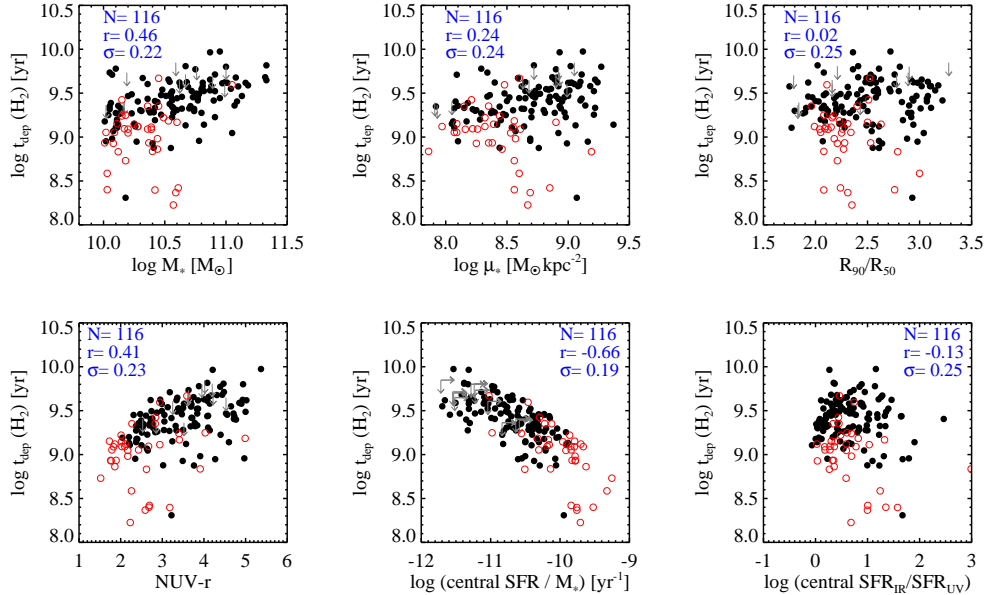}
    \caption{Molecular gas depletion time as a function of global galaxy properties. Black circles 
denote the representative sample of normal galaxies with secure CO, FUV, and IR measurements. 
Grey arrows denote the normal galaxies with secure CO detections but upper limits on their SFR estimates. 
Red circles represent the starburst galaxies. The linear relation is fitted 
to the normal galaxies with secure measurements  of both molecular gas and SFR. 
N, r, and $\sigma$ indicate the number of galaxies included in the
fit, the Pearson correlation coefficient, and the scatter about the best-fit relation.}
  \label{f4}
\end{center}
\end{figure*}

\begin{table*}   
\caption{Summary of the best fit linear relations between $t_{dep}$ and
a variety of galaxy parameters for SFR estimates
based on SED fitting (cols 4-7) and for SFR estimates based on
UV+22$\micron$ measurements (cols 8-11).  
The relations are parametrized as log t$_{dep}$ = m(x-x$_0$)+b}
\begin{center}
\begin{tabular}{cccccccccccc}
\hline\hline
            &                    &        & \multicolumn{4}{c}{SED fitting} & & \multicolumn{4}{c}{UV + IR}\\   
x parameter & Units	         & x$_0$  &  m       & b         &$\sigma$& r &  &    m  & b & $\sigma$  & r \\\hline
log $M_*$  &log $M_\odot$       & 10.7 &0.40$\pm$0.07&9.06$\pm$1.03&0.269   &0.48  &&0.33$\pm$0.06&9.44$\pm$0.02&0.22 & 0.46 \\
log $\mu_*$&log M$_{\odot}$kpc$^{-2}$ &8.7&0.40$\pm$0.07&8.99$\pm$0.88&0.275   &0.44  &&0.17$\pm$0.06&9.40$\pm$0.02&0.24   &0.24\\
R$_{90}$/R$_{50}$ & -               &2.5&0.37$\pm$0.07&8.99$\pm$0.22&0.290  &0.40  &&0.01$\pm$0.06&9.40$\pm$0.02&0.25 &0.02\\
NUV-r          & mag               &3.5&0.18$\pm$0.03&8.98$\pm$0.15&0.263   &0.49  &&0.12$\pm$0.03&9.41$\pm$0.02&0.23 &0.41\\
log SFR/M$_*$  & log yr$^{-1}$  & -10.40&-0.54$\pm$0.04&9.06$\pm$0.62&0.221  &-0.70 &&-0.37$\pm$0.04&9.30$\pm$0.02&0.19 &-0.66\\
SFR$_{IR}$/SFR$_{UV}$ & -       &  1   &       -       &   -           &  -  &  -   &&-0.06$\pm$0.05&9.38$\pm$0.03&0.25 &-0.13\\\hline
\end{tabular}
\end{center}
\label{tbl1}
\end{table*}

\subsection{HERACLES sample}  

\subsubsection{Calculation of molecular gas depletion time} 
We follow the procedures outlined in \citet{ler08} and \citet{big08} to 
calculate the molecular gas depletion time in 1-kpc $\times$ 1-kpc grid cells. 
Depletion time in our work is defined as $\rm \Sigma_{H_{2}}/ \Sigma_{SFR}$.

We briefly summarize our steps as follows. 
We first derive $\rm \Sigma_{H_{2}}$ in each 1-kpc grid cell from the reduced 
HERACLES CO maps adopting the Galactic $\rm \alpha_{CO}$ value,
4.35 M$_{\sun}$ pc$^{-2}$ (K km s$^{-1}$)$^{-1}$.
To derive the $\rm \Sigma_{SFR}$, 
we apply the colour-based masks described in \citet{mun} 
to remove stars and background galaxies for the FUV and 
24$\mu$m images. We check the images again by visual inspection and 
mask the stars by hand when necessary. 
We correct the FUV images for Galactic extinction  using the maps of \citet{sch}.
All the images are convolved to  $\sim$13\arcsec HERACLES resolution  using 
the kernels provided in \citet{ani}.
\citet{ler12} pointed out  that  cirrus emission from old stellar 
populations could contribute up to $\sim20\%$ of the total IR luminosity.  
To remove the cirrus contribution,
we apply the method suggested in sec 8.2 of \citet{ler12}.
{\bf With gas, 24 $\micron$, and FUV data in hand,} 
they recommended that a first-order cirrus emission
correction could be made using the total  gas surface density, i.e. 
the sum of $\rm \Sigma_{H_{2}}$ and $\rm \Sigma_{HI}$,
where $\rm \Sigma_{HI}$ is obtained from THINGS data.
After we remove this cirrus emission from the 24 $\micron$ emission following 
equation (15) in \citet{ler12}, we then adopt their updated calibration 
coefficient to calculate the SFRs from the linear combination of the FUV and 
24 $\micron$ luminosities.  

We exclude bins with $\Sigma$$\rm _{SFR}$ $<$ 10$^{-3}$ M$_{\sun}$ yr$^{-1}$ kpc$^{-2}$
since the calibration becomes poor at very low SFR \citep{ler12}.
We also discard the bins with $\rm \Sigma_{H_{2}}$ $<$  3 M$_{\sun}$ pc$^{-2}$,
which is the sensitivity limit for the HERACLES CO maps. 

\subsubsection{Derivation of local parameters} 
{\bf We also derive the local parameters, M$_{*}$, stellar mass surface 
density ($\mu_{*}$),  NUV-r colour, and specific SFR (sSFR) 
for galaxies in the HERACLES sample.} 
The local parameters are calculated within the 1-kpc scale grids. 
We measure the SDSS {\it ugriz} magnitudes and GALEX FUV, NUV magnitudes 
after we remove stars and correct for Galactic foreground extinction using the 
methods described above. We use the SED-fitting method in \citet{wan}
to derive the stellar mass. {\bf  The stellar mass surface density is the stellar mass
divided by the physical area of each grid cell , in units of  M$_{\sun}$ kpc$^{-2}$. }
The NUV-r colour is defined as the SDSS $r$-band minus the GALEX NUV magnitude.
The parameter, sSFR, is the SFR divided by the stellar mass within each grid cell.
We note that M$_{*}$ = $\mu_{*}$ for 1 kpc square grid cells.

{\bf Because \citet{ler13} estimated the stellar mass using 3.6$\micron$ fluxes,
different from our SED-fitting method, one might worry that this would play a 
role in our results. As \citet{ler13} provides the global values of M$_*$ 
estimated from 3.6$\micron$ fluxes for the  HERACLES sample, we compare our values 
with theirs. The linear coefficient is 0.9 with scatter 0.11; we do not see a 
significant systematic difference.}

\begin{figure*} 
\begin{center}
 \includegraphics[scale=0.67]{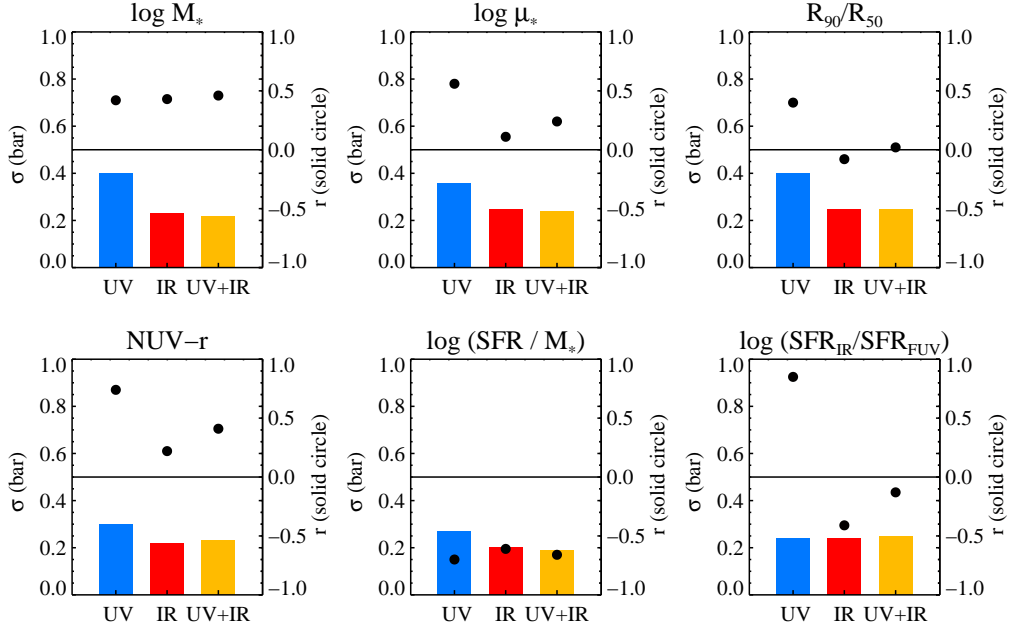}
    \caption{The scatter ($\sigma$) and the correlation coefficient (r) of the fitted linear relations between
             depletion time and galaxy parameters based on 3 different SFR tracers: 
             UV+IR, IR, FUV. The bar plots show the value of the scatter as indicated on the left y-axis;
             the solid circles denote the correlation coefficients as indicated on the right y-axis.}
  \label{f5}
\end{center}
\end{figure*}

\section{Results} 

In this section, we present the results of our new derivation of depletion time  
and we compare  with the results of \citet{sanb}. 
We also examine the systematic effects that cause the discrepancy
between our results and those of the 2011 study.  
Finally, we show that we can identify a single  global galaxy  parameter 
that drives variations in  depletion time among different galaxies.

As in Saintonge et al (20011b), the global galaxy parameters that we investigate are  M$_*$, 
$\mu_*$, R$_{90}$/R$_{50}$, NUV-r, and sSFR. In addition,  we consider the ratio
of the infrared-to-UV flux, IR/UV.
The stellar mass, M$_*$, is an approximate tracer of the global potential well 
depth of the galaxy.  
The stellar surface density, $\mu_*$, and the concentration index, 
R$_{90}$/R$_{50}$, are structural parameters.
The concentration index is tightly correlated with the bulge-to-disc ratio.
The stellar surface mass density is a convenient way of scaling galaxy size and stellar 
mass and encodes physical information about angular momentum conservation/loss during the 
process of galaxy formation. The global NUV-r colour is a measure of the ratio of young 
stars that are not obscured by dust to older stars.  
The sSFR parameter is the ratio of the current star formation rate, corrected
for dust extinction, to the total stellar mass. 
For consistency, we derive sSFR within the same 22\arcsec area of the galaxy where we  
measure the molecular gas mass.
The ratio of the total IR to FUV luminosity is a robust 
estimate of the  dust attenuation (e.g., \citealt{gor}; \citealt{bel}; \citealt{hao}). 
Although we do not have enough multi-wavelength data to estimate the total IR luminosity, 
the 24\micron\ luminosity has been shown to correlate linearly with the 
total IR luminosity \citep{wu,cal,rie}.  
Thus, we use the ratio of the SFR$_{22\micron}$ to SFR$\rm_{FUV}$ (IR/UV) as our indicator of 
dust attenuation.

\subsection{Depletion time based on our new SFR estimates}

We plot depletion time results based on our new SFR estimates as a 
function of global physical properties of galaxies in Fig.\ref{f4}. 
The black circles show galaxies in the representative sample with 
secure CO, FUV and 22\micron\ detections. The grey arrows show 
galaxies with secure CO detections but with  SFR upper limits. 
An extra set of galaxies with high values of SFR/$M_*$ were observed at the end
of the COLD GASS program in order to extend the dynamic range in star formation
probed by the observations. We will refer to these galaxies as ``starburst''
systems and we indicate them as red symbols on the figure. 
Some of these galaxies have enhanced infrared luminosity
and dust temperature and we thus use the  ``ULIRG'' value  of $\rm \alpha_{co}$  
(1 M$_{\sun}$ pc$^{-2}$ (K km s$^{-1}$)$^{-1}$) to derive their molecular
gas masses.
We note that some of these starburst galaxies are interacting/merging systems and
it is inevitable to include emission from their companions  when we calculate their global properties. 
To minimize any confusion, we visually identify eight interacting/merging systems and exclude them in
the following analyses and discussions.
We perform linear fits to galaxies from the representative sample only and the    
number of galaxies in the sample (N), the Pearson 
linear correlation coefficient (r), and the scatter ($\sigma$)  
are indicated on the figure. 

Our new linear relations are compared with those derived by \citet{sanb}  
in Table 1. Three main changes can be seen:   
\begin {enumerate}
\item The normalization $b$ changes. The resulting depletion times are a factor of two longer than quoted
in \citet{sanb} and now agree very well with HERACLES results (see later). The change
in normalization occurs because the apertures within which we measure the molecular
gas and the SFR are now well-matched.    
\item The scatter $\sigma$ in the relations between depletion time and global galaxy parameters
is significantly reduced. This occurs for all the relations  and is a consequence
of our improved star formation rate estimates, which do not have to be corrected for
dust in a statistical fashion.
\item The slopes of the relations between depletion time and the structural parameters, 
$\mu_*$ and concentration index $R_{90}/R_{50}$, are significantly flatter than before.
In particular, there is now no correlation at all between depletion time and galaxy
bulge-to-disk ratio, whereas \citet{sanb} found a very significant correlation. 
\end {enumerate}
There are still strong correlations between the depletion time and stellar 
mass, NUV-r color and sSFR, with only small changes in slope from previously.
We note that the depletion time is not dependent on the IR/UV ratio.
In the next section, we investigate  the systematic effects in the dust 
extinction properties of the galaxies, which are the main cause of the discrepancy
between our work and that of \citet{sanb}.

\begin{figure*} 
\begin{center}
 \includegraphics[scale=0.52]{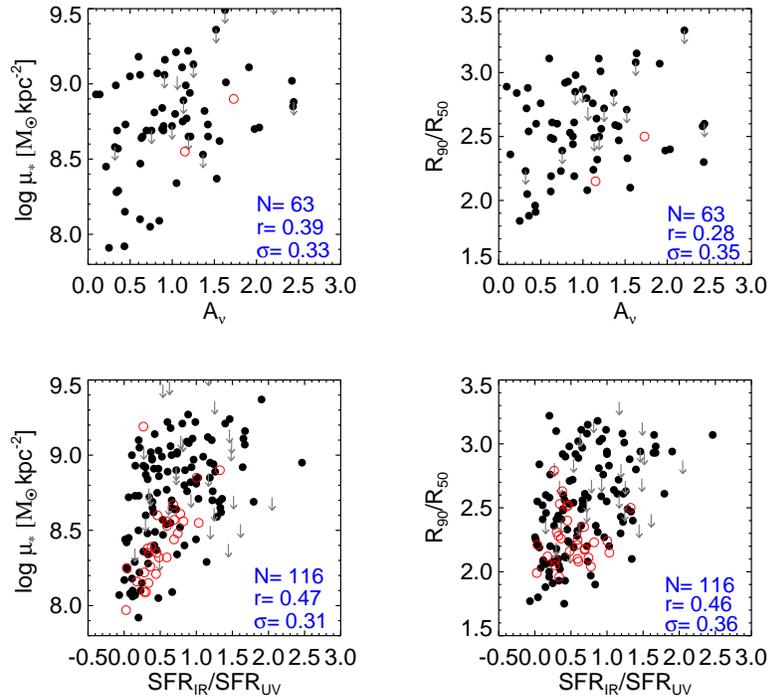}
    \caption{Stellar surface density and concentration index as functions of A$\nu$ and IR/UV. }
  \label{f6}
\end{center}
\end{figure*}

\begin{figure*} 
\begin{center}
 \includegraphics[scale=0.52]{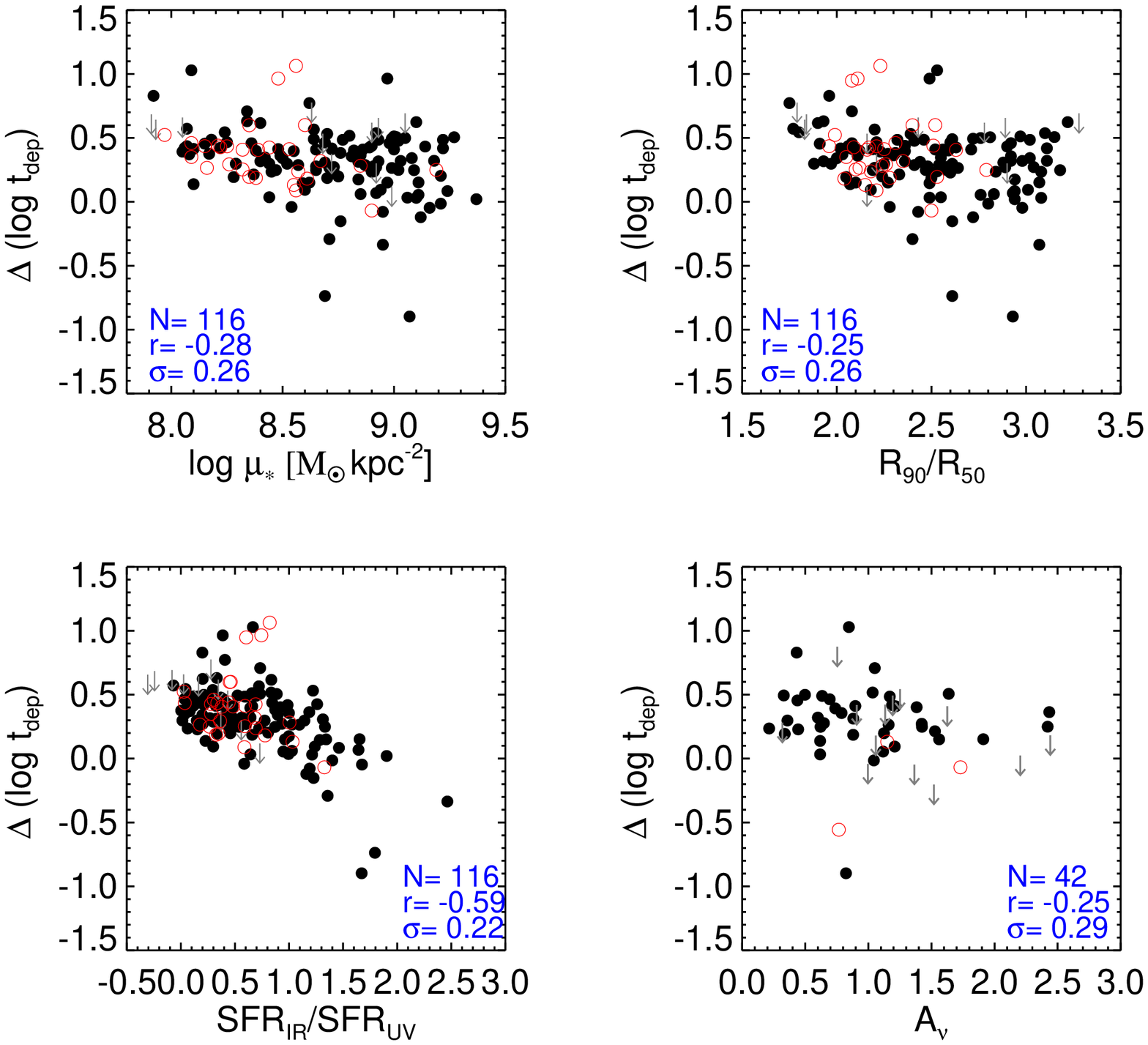}
    \caption{Difference in the depletion time , $\Delta$ (log t$_{dep}$), measured in \citet{sanb} and 
             our work as functions of stellar surface density, concentration, IR/UV ratio and dust extinction. }
  \label{f7}
\end{center}
\end{figure*}

\subsection{Systematic effects caused by dust attenuation}

In Fig. \ref{f5}, we illustrate how the correlation coefficients $r$ and the 
scatter $\sigma$ about the best-fit linear relation between molecular gas depletion time
and galaxy properties change when we derive star formation rate purely from the
UV flux (SFR$\rm_{UV}$), purely from the IR flux (SFR$\rm_{IR}$) and from the sum of the 
two (SFR$\rm_{UV+IR}$). The bar 
plots show the value of $\sigma$, as indicated on the left-hand axis; the solid circles 
show the correlation coefficients, as indicated on the right-hand axis. We first observe that 
the scatter tends to be significantly larger for the UV-based relations, and of 
comparable magnitude for IR and UV+IR-based relations. 
The strength of the correlations with the structural parameters $\mu_*$ and R$_{90}$/R$_{50}$
is only strong for the UV-based relations, and largely disappears for the  SFR$\rm_{IR}$
and SFR$\rm_{UV+IR}$ relations. In contrast, the strength of the
correlations with stellar mass $M_*$ and specific star formation rate SFR/M$_*$
are the same for  SFR$\rm_{UV}$,  SFR$\rm_{IR}$ and  SFR$\rm_{UV+IR}$.
This supports our conjecture that systematics in IR/UV ratio (or dust extinction) are the
main reason for the changes between the results presented in this paper and those
in \citet{sanb}. 

As both t$_{dep}$ and IR/UV include the UV flux, it is worth using an independent
estimate of dust extinction to investigate the effect of dust systematics on our results.
A sub-sample of the COLD GASS galaxies have 
long-slit spectra along their major axes \citep{mor}, from which we can calculate the
extinction A$_{\nu}$ from the Balmer decrement, L(H$_\alpha$)/L(H$_\beta$).
The spectra were spatially binned outward from the galaxy center to 
ensure an adequate $\rm S/N$ in every bin.  We exclude the bins 
with low $\rm S/N$ for the Balmer lines and weight the  bins with reliable measurements 
by the stellar mass contained within the annulus spanned by
the bin boundaries, to get a representative A$_{\nu}$ value for the whole galaxy.

In Fig. \ref{f6}, we plot $\mu_{*}$ and R$_{90}$/R$_{50}$ as a function of 
$A_{\nu}$ and the IR/UV ratio. These structural parameters
correlate strongly with both  $A_{\nu}$ and IR/UV. In order to ascertain whether
these correlations are responsible for  differences in our depletion time results  
with respect to those of Saintonge et al (2011b), 
we calculate the difference between the two depletion
time estimates  $\Delta$ log (t$_{dep}$)= log t$_{dep}$ (this paper) 
$-$ log t$_{dep}$ \citep{sanb}. In Fig. \ref{f7},  $\Delta$ log (t$_{dep}$) is plotted against 
$\mu_{*}$, R$_{90}$/R$_{50}$, IR/UV, and $A_{\nu}$.
As can be seen, our depletion time estimates become shorter with respect to those in \citet{sanb} 
when galaxies have  higher stellar surface densities and larger concentration indices. 
The difference $\Delta$ log (t$_{dep}$) is also correlated with 
IR/UV and $A_{\nu}$.

\begin{figure*} 
\begin{center}
 \includegraphics[scale=0.55]{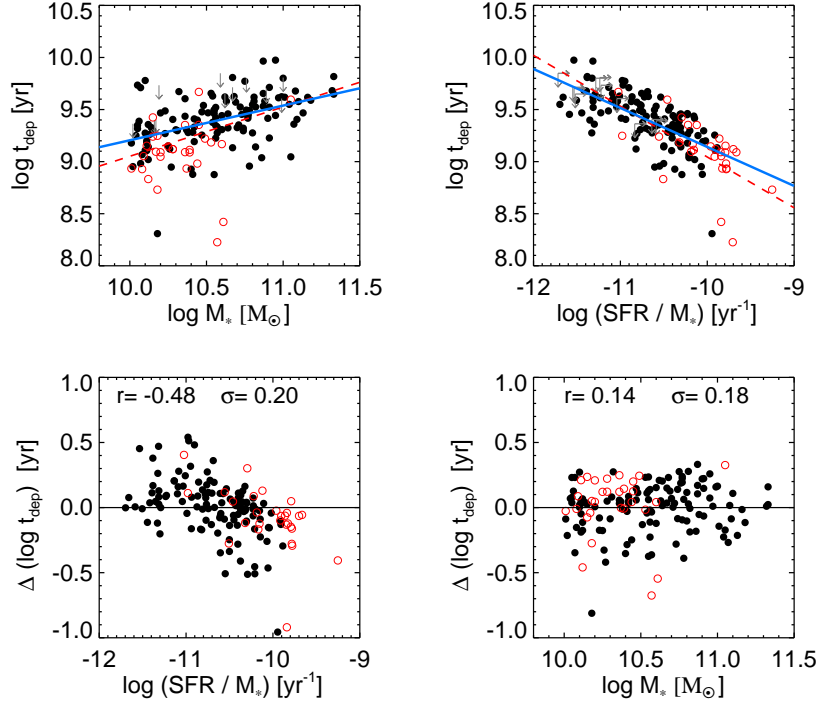}
    \caption{Upper panels: molecular gas depletion time is plotted
as a function of $M_*$ and SFR/$M_*$. Blue lines show the  best-fit linear relation for galaxies in
the representative sample, while the red lines show the best-fit linear relation for the sample
that includes the starburst systems.  
Bottom left : The residuals from the best fit $t_{dep}- M_*$ relation are plotted as a function
of SFR/$M_*$. Bottom right:  The residuals from the best fit $t_{dep}- SFR/M_*$ relation are 
plotted as a function of M$_*$.}
  \label{f8}
\end{center}
\end{figure*}

\begin{figure*} 
\begin{center}
 \includegraphics[scale=0.55]{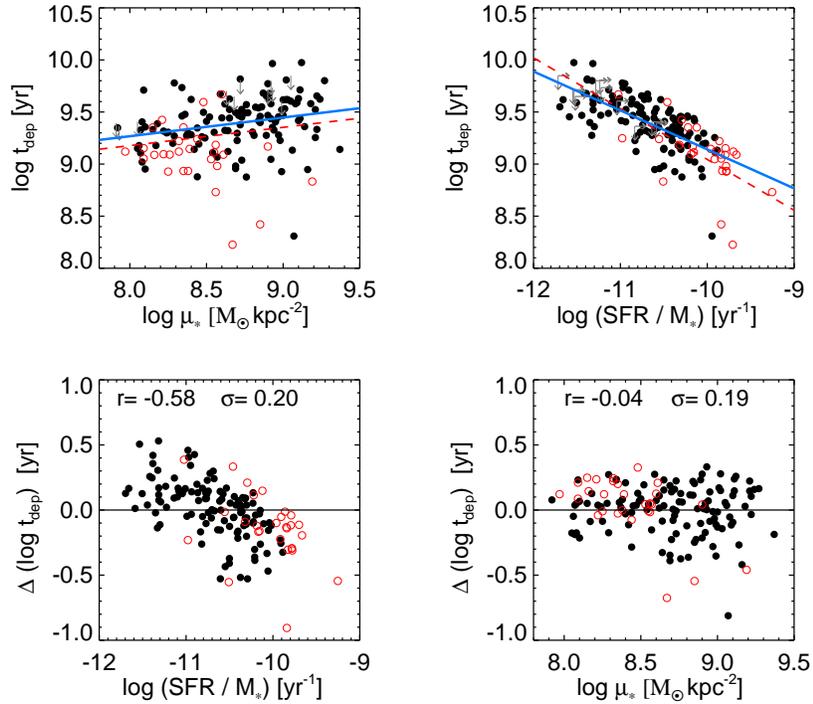}
    \caption{As in Figure 8, except that $M_*$ has been replaced by $\mu_*$} 
  \label{f9}
\end{center}
\end{figure*}

\begin{figure*} 
\begin{center}
 \includegraphics[scale=0.6]{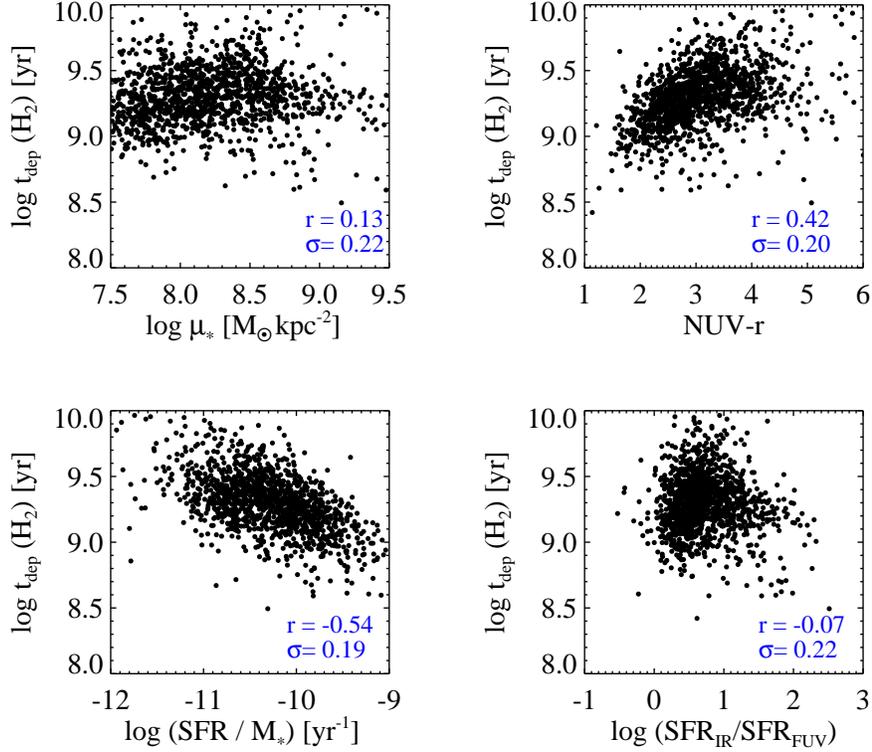}
    \caption{Molecular gas depletion times evaluated within 1kpc$^2$ area grid cells as functions of 
   stellar surface density, NUV-r color, sSFR and IR/UV for galaxies from the HERACLES sample.}
  \label{f10}
\end{center}
\end{figure*}

\begin{table*}   
\caption{Summary of the best-fit linear relations between
$t_{dep}$ and local parameters for grid scale data from the HERACLES sample. 
The relations are parametrized as log t$_{dep}$ = m(x-x$_0$)+b}
\begin{center}
\begin{tabular}{cccccccc}
\hline\hline
x parameter & Units	          & x$_0$  &  m       & b         &$\sigma$& r &   \\\hline
log $\mu_*$&log M$_{\odot}$kpc$^{-2}$&8.7     &0.06$\pm$0.01  &9.33$\pm$0.01&0.22   &0.13\\
NUV-r          & mag              &3.5     &0.12$\pm$0.0 1 &9.35$\pm$0.01&0.20   &0.42 \\
log SFR/M$_*$  & log yr$^{-1}$    & -10.40 &-0.24$\pm$0.01 &9.32$\pm$0.01&0.19  &-0.54\\
SFR$_{IR}$/SFR$_{UV}$ & -         &  1     & -0.04$\pm$0.01&9.29$\pm$0.01&0.22&-0.07 \\\hline
\end{tabular}
\end{center}
\label{tbl2}
\end{table*}

\subsection{Primary or Induced Correlation?}
Our results show that molecular gas depletion time is strongly correlated with 
M$_*$, NUV-r and sSFR, and is more weakly correlated with $\mu_*$.  
One question is whether these correlations are independent of each other.
Almost all global galaxy parameters correlate strongly
with stellar mass,  so one might expect
the molecular gas depletion time to correlate with $M_*$ even if the 
{\em primary} correlation is  with some other parameter. In this section,
we attempt to separate primary and induced correlations through
analysis of residuals.  
We note that NUV-r and sSFR have similar physical meaning in that they both serve as
an indicator of the ratio of young-to-old stars in the galaxy, so for simplicity
we only consider sSFR, which does not depend on dust extinction.

We fit a linear relation to the t$_{dep}$ versus $M_*$ relation and 
ask whether the residuals of t$_{dep}$ from this relation are correlated with sSFR.   
Similarly, we also fit a linear function to 
the t$_{dep}$--sSFR relation and examine if the residuals correlate 
with M$_*$. The results are shown in Fig. \ref{f8}. 
The blue line shows the fit to the representative sample, while the red line shows  
the fit to the combination of  normal and starburst galaxies.
The residuals are measured with respect to the blue line.         

As seen in the lower panel of Fig. \ref{f8}, the residuals from the  
t$_{dep}$--M$_*$ relation correlate strongly with sSFR, while the residuals from the 
t$_{dep}$--sSFR relation correlate very weakly  with M$_*$.
We note that there is one  galaxy with particularly low depletion time (0.1 Gyr)
and with a central sSFR $\sim$10$^{-9.9}$ yr$^{-1}$,
which causes most of this weak correlation.
In the MPA-JHU catalog, this galaxy has been classified as a galaxy with an AGN in the center. 
If this galaxy is excluded, there is no longer any significant correlation between
the t$_{dep}$ residuals and stellar mass.  

We carry out the same exercise for the relation between t$_{dep}$ and $\mu_*$. 
The results are plotted in  Fig. \ref{f9}. The residuals of the 
t$_{dep}$--$\mu_*$ relation are strongly correlated with sSFR, 
with a significance r =$-0.58$. By contrast, there is no 
relation between $\mu_*$ and the residuals of the t$_{dep}$--sSFR relation.
We again conclude that the primary relation is the one between t$_{dep}$ and sSFR.

{\bf The trend of t$_{dep}$ with sSFR on global scale can be also seen in the
results based on a fixed $\alpha_{CO}$ from \citet{ler13}. In their Table 6, the 
rank coefficient (1st column) for t$_{dep}$--sSFR relation is $-0.38$ for all 
sample and $-0.27$ for the sample with log M$_*$ $>$ 10. This indicates there is  
a dependence of t$_{dep}$ on sSFR though this dependence is weaker than our finding. 
This might be due to differences in the sample size and the range of the 
galaxy properties, as discussed in Sec 1. }

\subsection{Comparison with results for 1 kpc grids} 

In order to examine whether the trends we find in the COLD GASS sample are 
similar to those found on smaller (1 kpc) scales, we utilize data   
from the HERACLES project. We derive depletion time on grids with 1 kpc$^2$ area   
and study the dependence on local physical properties, such as  
stellar surface density, NUV-r color, sSFR, and  IR/UV ratio. 
As discussed in section 3.2, {\bf bins with low SFR surface densities are excluded} and     
we also remove the IR cirrus emission from old stellar 
populations following the procedures outlined in \citet{ler12}.

We plot the molecular gas depletion time versus stellar surface mass density,
NUV-r color, sSFR, and  IR/UV ratio in Fig. \ref{f10}. 
As in the COLD GASS sample, the depletion time on 1 kpc grid scales in the HERACLES sample 
is  strongly correlated with NUV-r and sSFR, but does not
show significant dependence on the local stellar mass density or  
the IR/UV ratio. 

To quantify the strength of the correlations on local scales, we also fit linear 
relations to the HERACLES data and summarize the parameters of the best-fit relations in Table 2. 
Generally, the slopes and the correlation coefficients are somewhat lower for the grid 
measurements compared to the  COLD GASS sample.
The slope and the significance of the t$_{dep}$--sSFR correlation on grid scales 
are $-0.24$ and $-0.54$, compared to  $-0.37$ 
and $-0.66$ on global scales.

\begin{figure} 
\begin{center}
 \includegraphics[scale=0.57]{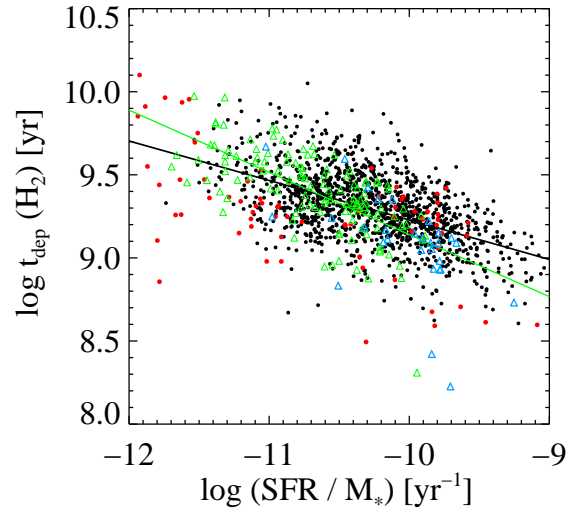}
    \caption{ Comparison of the global t$_{dep}$--sSFR relation for COLD GASS galaxies
             and for measurements on 1 kpc grid scales for galaxies in the HERACLES sample. 
             Green and blue triangles denote the representative and starburst samples from the
             COLD GASS survey. 
             Black and red points denote the grid measurements in {\em disc} and {\em bulge} regions
             of galaxies from the HERACLES sample.
            Linear relations 
            are fit to both the COLD GASS representative sample (green line) and the HERACLES 
            data (black line).}
  \label{f11}
\end{center}
\end{figure}

We plot the t$_{dep}$--sSFR relations for grid and global scales, as well as the best-fit
linear relations,  in Fig. \ref{f11}. 
We separate the grid cells into those located in the bulge-dominated and  disk-dominated
regions of galaxies as follows.
We apply the GALFIT code \citep{pen} to the SDSS r-band images to decompose galaxies into 
bulges and disks. To fit the luminosity profile of the galaxies, the code assumes a two-component model, 
Sersic bulge plus Sersic disk profile, where the Sersic $n$ values for the bulge and 
disk are allowed to vary from 1.5 to 4 and from 0.8 to 1.2, respectively. 

Two main conclusions emerge from the results shown in Fig. \ref{f11}:
a) the global t$_{dep}$--sSFR relation for the COLD GAS galaxies overlaps that for
the {\em disk grid cells} quite well. b) The number of grid cells
in bulge-dominated regions is quite low. Some of these bulge grid cells 
do fall below the fitting relation.
This is consistent with the results in \citet{ler13}, where shorter
molecular gas depletion time were found in the nuclear nuclear regions of some galaxies. 
Nevertheless, it is quite clear from Fig. \ref{f11} that the main t$_{dep}$--sSFR
correlation is driven by disk-dominated regions of the galaxy.

Because the SFR estimate is included in the calculation of both t$_{dep}$ and sSFR, 
one might worry that the correlation between t$_{dep}$ and sSFR might be induced 
rather than a real correlation. This issue was addressed in detail in \citet{sanb}.
It was demonstrated that molecular gas depletion time correlated very well with
4000 \AA\ break strength measured from the SDSS fiber spectrum, which provides
an independent estimate of the ratio of young-to-old stars in the galaxy. In addition, these
authors carried out Monte Carlo simulations where they showed that the observed relation was
too strong to be ``induced'' by scatter arising from measurement errors.

\begin{figure} 
\begin{center}
 \includegraphics[scale=0.5]{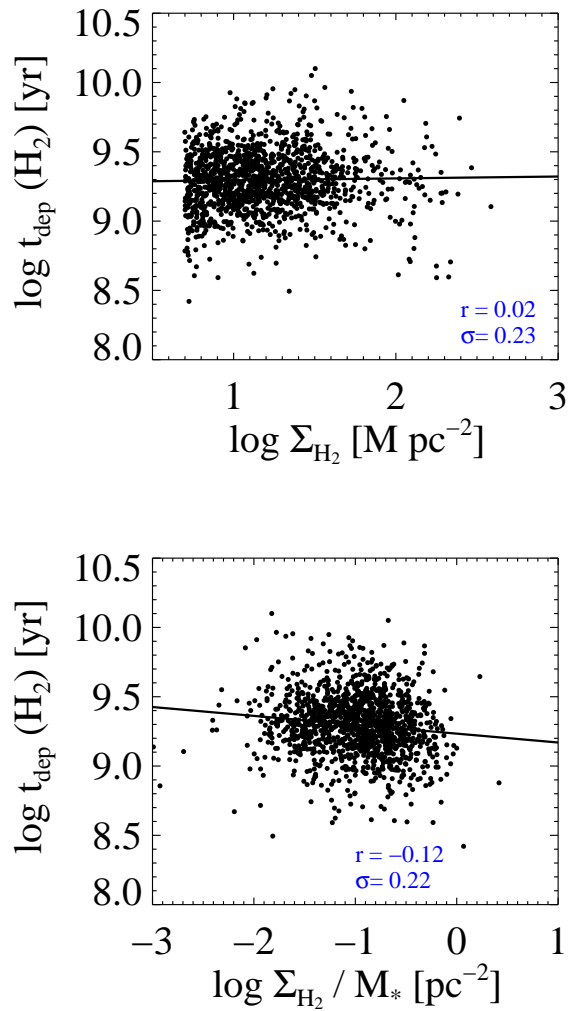}
    \caption{Depletion time as a function of molecular gas surface density and molecular gas 
surface density per unit stellar mass for the HERACLES grid measurements.}
  \label{f12}
\end{center}
\end{figure}

In Fig. \ref{f12}, we investigate whether the depletion time has dependence on the molecular
gas surface density, $\rm \Sigma_{H_2}$ and the molecular gas surface density
scaled by the stellar mass within the grid cell, $\rm \Sigma_{H_2}/M_*$. 
As can be seen, there is no dependence of  t$_{dep}$ on 
$\rm \Sigma_{H_2}$. However, when $\rm \Sigma_{H_2}$ is scaled by the stellar mass
measured in the grid cell, a mild dependence on $\rm \Sigma_{H_2}/M_*$, 
with the linear coefficient r = $-0.11$, is found. {\bf This result might indicate  
that the {\em local density of evolved stars has an effect on
the molecular gas depletion time}. }

We conclude that molecular gas is depleted most quickly in regions of the galaxy where the 
current star formation rate is high and the density of already-formed stars is low. This 
may be true in spiral arm regions -- this is a hypothesis that we will investigate in a 
future paper.

\section{Summary and Discussion}

In this paper, we re-analyze the relations between global molecular gas depletion time
and a variety of galaxy parameters for nearby galaxies from the COLD GASS
survey with stellar masses in the range
10$^{10}$ -- 10$^{11.5}$ M$_{\sun}$ and redshifts in the range 0.02.-- 0.05.
The molecular gas mass is estimated from the CO(J=1-0) line measurements  
and our updated estimates of  star formations use the combination of GALEX 
FUV and WISE 22 $\micron$ data. In agreement with \citet{sanb},
we find that the molecular gas depletion 
time depends strongly on  galaxy stellar mass,  NUV-r colour 
and sSFR. Our results differ from those in \citet{sanb}
in that we find that the dependences of the depletion time on galaxy
structural parameters such as  stellar surface density and concentration index, 
are weak or absent. We demonstrate  that the differences   
with the \citet{sanb} analysis  arise from the fact that dust extinction
as measured by the ratio of 22 micron  to far-UV
flux in the galaxy, correlates strongly with $\mu_*$ and concentration.  
We further demonstrate that the dependence of t$_{dep}$ on M$_*$ is  
actually driven by the primary correlation between t$_{dep}$ and sSFR. 

We compare our results with molecular gas depletion time estimates on   
1-kpc scales using publicly available data from the HERACLES survey. 
We find remarkably good agreement with our global t$_{dep}$ versus sSFR relation.  
On sub-galactic scales, we are able to ascertain that t$_{dep}$ is not 
correlated with $\Sigma_{\rm H_2}$, but is weakly correlated with
$\Sigma_{\rm H_2}/M_*$, indicating that the presence of old stars has an effect 
on the ability of molecular clouds to form new stars.

We note that the parameter, sSFR, is the ratio of current SFR to the stellar mass 
built up by the past star formation. The strong correlation between t$_{dep}$ and sSFR, 
extending over a factor of 10 in t$_{dep}$ in our sample, leads to the 
inference that the molecular gas depletion time is dependent on  the star 
formation history of the galaxies. Those galaxies with high current-to-past averaged 
star formation activity, will drain their molecular gas reservoir sooner.


The influence of metallicity on  $\rm \alpha_{CO}$ value is always a concern 
whenever the star-formation law is studied. 
One might question whether our assumption of a fixed $\rm \alpha_{CO}$ value is appropriate.  
\citet{san} studied variations in  $\rm \alpha_{CO}$ using a subset 
of 26 galaxies from the HERACLES sample. They 
showed  that metallicity effects on  $\rm \alpha_{co}$ are very weak over  
the metallicity range spanned by the HERACLES disc galaxies.
$\rm \alpha_{co}$ has been shown to depart significantly from the typical Galactic value
only below a metallicity $\sim$ 1/3 -- 1/2 solar (e.g., Glover \& Mac Low 2011; 
Leroy et al. 2011; Bolatto et al. 2013). 
We would also expect to see a dependence of gas depletion time on  
the dust content of the galaxy, if our results were simply a consequence of variations in   
$\rm \alpha_{CO}$. From Fig. \ref{f4} and \ref{f10},
we know that the depletion time  on both global and local scales 
does not show any correlation with IR/UV ratio.

\citet{san} did find that $\rm \alpha_{CO}$ is dependent on other 
galaxy parameters, such as average radiation field intensity, 
PAH fraction, stellar mass surface density, SFR surface density and dust mass 
surface density. The $\rm \alpha_{CO}$ value appears to decrease as the stellar mass and SFR
surface densities increase. We note that these two effects cancel  when we derive 
sSFR. We conclude that it is unlikely that the small variations in $\rm \alpha_{CO}$
found by \citet{san} explain our observed relation between t$_{dep}$ and sSFR.

In future work, we will investigate whether the strong t$_{dep}$--sSFR is connected to the structure of 
the interstellar medium, such as the presence of spiral arms. Gas clouds are compressed in  spiral arm regions
and one might hypothesize that galaxies with high SFR/M$_{*}$ 
might have more arm structures than those with low SFR/M$_{*}$, leading to   
the observed variation in molecular depletion time scales in disks.

\section*{Acknowledgments}
We thank Am\'elie Saintonge, Jing Wang, Richard D'Souza, 
Jarle Brinchmann, Sambit Roychowdhury, Li Shao and Frank Bigiel for helpful discussions.

GALEX is a NASA Small Explorer, launched in 2003 April, developed in cooperation with the Centre National d¡¯Etudes Spatiales of France and the Korean Ministry of Science and Technology.

This publication makes use of data products from the Wide-field Infrared Survey Explorer, which is a joint project of the University of California, Los Angeles, and the Jet Propulsion Laboratory/California Institute of Technology, funded by the National Aeronautics and Space Administration. 

Funding for the SDSS and SDSS-II has been provided by the Alfred P. Sloan Foundation, the Participating Institutions, the National Science Foundation, the US Department of Energy, the National Aeronautics and Space Administration, the Japanese Monbukagakusho, the Max Planck Society, and the Higher Education Funding Council for England. The SDSS is managed by the Astrophysical Research Consortium for the Participating Institutions. The Participating Institutions are the American Museum of Natural History, Astrophysical Institute Potsdam, University of Basel, Cambridge University, Case Western Reserve University, University of Chicago, Drexel University, Fermi National Accelerator Laboratory, the Institute for Advanced Study, the Japan Participation Group, Johns Hopkins University, the Joint Institute for Nuclear Astrophysics, the Kavli Institute for Particle Astrophysics and Cosmology, the Korean Scientist Group, The Chinese Academy of Sciences (LAMOST), the Leibniz Institute for Astrophysics, Los Alamos National Laboratory, the Max-Planck-Institute for Astronomy (MPIA), the Max-Planck-Institute for Astrophysics (MPA), New Mexico State University, Ohio State University, University of Pittsburgh, University of Portsmouth, Princeton University, the US Naval Observatory and the University of Washington.

\label{lastpage}

\end{document}